# Selective nonthermal melting in phlogopite under ultrafast energy deposition

Nikita Medvedev[1,2,*]

*1) Institute of Physics, Czech Academy of Sciences, Na Slovance 1999/2, 182 00 Prague 8, Czech Republic*
*2) Institute of Plasma Physics, Czech Academy of Sciences, Za Slovankou 3, 182 00 Prague 8, Czech Republic*

## Abstract

Phlogopite is a complex magnesium-rich mineral from the dark mica group, $KMg_3(AlSi_3O_{10})(OH)_2$. Its response to ultrafast excitation of its electronic system is studied using a hybrid model that combines tight-binding molecular dynamics with transport Monte Carlo and the Boltzmann equation. Simulations predict that at the deposited dose of ~0.17 eV/atom (electronic temperature $T_e$~11,000 K), the first hydrogens start to migrate in the otherwise preserved lattice, transiently turning mica into a superionic state. At the dose of ~0.4 eV/atom ($T_e$~13,000 K), Mg atoms start to diffuse like a liquid within stable sublattices of other elements, suggesting a superionic-superionic phase transition. At a dose of approximately 0.5 eV/atom ($T_e$~14,000 K), the entire atomic lattice destabilizes, disordering on picosecond timescale. It is accompanied by the formation of defect energy levels inside the bandgap. At the doses ~0.9 eV/atom ($T_e$~16,000 K), the bandgap completely collapses, turning the material metallic (electronically conducting). At even higher doses, nonthermal acceleration of atoms heats the atomic system at ultrafast timescales; K and O elements are most affected, accelerating within a few tens of femtoseconds.

## I. Introduction

Phlogopite ($KMg_3(AlSi_3O_{10})(OH)_2$), an example of dark mica, is a mineral with a monoclinic structure that forms flexible and elastic layers or flakes. The trioctahedral layers are connected by potassium sites; magnesium fills octahedral positions, whereas the tetrahedral sheets are occupied by a mix of silicon and aluminum atoms [1]. It has low resistance to ionizing radiation (including gamma or beta rays, or alpha particles and heavy ions) under conventional low-dose-rate conditions [2–4]. Phlogopite's abundance in Earth's crust instigated its various industrial and research applications, such as electrical and heat insulation, plastic and rubber reinforcement, use in construction materials, paints and coatings, as well as radiation-related applications [5–8].

---

[*] Corresponding author: email: nikita.medvedev@fzu.cz, ORCID: 0000-0003-0491-1090

Phlogopite susceptibility to ionizing radiation is used in a geological dating method, registering recoil tracks created by the natural α-decay of U, Th, and their fission products [4]. A swift ion (decay product) leaves a track of structurally modified material, a few nanometers in diameter and some microns to millimeters in length [4,9]. Such tracks form via a sequence of processes, starting with the excitation of electrons by the propagating ion, and eventually converting into observable atomic disorder [9,10].

Phlogopite's layered structure is similar to the matrix of clay, which motivated research on its electron and gamma radiation, to study the candidate materials for radionuclide waste storage [2,3]. Clay is considered a backfill material to prevent radionuclide migration. Understanding the governing mechanism of radiation damage in layered geological materials thus triggered research in phlogopite [2].

Recently, exfoliation of the natural dark mica into ultrathin layers or flakes has found applications in dielectric layers for 2d-optoelectronics [11]. Application of such devices implies exposure to electromagnetic radiation. Additionally, nano-electronic production often involves laser patterning of materials to tailor their properties [12].

The common effect in all these irradiation scenarios is that they are all initiated by the excitation of the electronic ensemble of the target [10,13,14]. The electronic system, driven out of equilibrium, undergoes the electron cascades of secondary ionizations, thermalizing and exchanging energy with the atomic system (the electron-phonon coupling) [13]. Atomic heating, overcoming the melting point, may lead to phase transitions, forming new material states [15,16]. At the same time, high electronic excitation induces a modification of the interatomic potential, which may destabilize the lattice and lead to disruption of atomic bonds (also known as nonthermal melting) [17,18].

Quantitative understanding of the various effects leading to final material modifications is required for practical applications. It must necessarily include the nonequilibrium effects in both electronic and atomic systems, phase transitions, and chemical bond evolution, describing the physics and chemistry of the transient states outside of the materials' phase diagram. Ab-initio methods, such as density-functional method, are limited to small simulation boxes, and usually do not include nonadiabatic electron-ion coupling [19,20]. Classical molecular dynamics (MD) simulations, on the other hand, allow for a large system treatment, but do not include electronic dynamics and nonthermal effects (changes in the interatomic potential due to electronic excitation) [21,22].

To study the phlogopite's response to radiation and various damage induced by irradiation, the XTANT-3 code is used here [23]. It combines a few approaches into a unified model with feedback: tight-binding molecular dynamics, transport Monte Carlo, and Boltzmann collision integral methods, delivering a state-of-the-art simulation method. This method enables to study electronic and atomic dynamics, modelling intertwined effects of the thermal, nonthermal, and nonequilibrium kinetics in the irradiated phlogopite. Simultaneously, it is capable of modeling sufficiently large simulation boxes to capture effects of a complex material.

## II. Model

The hybrid code XTANT-3 includes the following approaches to trace the effects of irradiation: the transport Monte Carlo method to describe irradiation and kinetics of fast electrons and deep-shell holes; the Boltzmann equation to trace the slow electrons populating the valence and the bottom of the conduction band; and the tight-binding molecular dynamics propagating the atomic trajectories on the evolving potential energy surface [23]. All the numerical details of the simulation can be found in the XTANT-3 manual [23], here we briefly outline the physical processes and models describing them.

The electronic excitation, induced by the photoabsorption, the following nonequilibrium electron cascades, and the Auger decays of core holes are modelled with the event-by-event individual particle transport Monte Carlo method [10]. The EPICS2025 database is used to extract the photoabsorption cross sections, the atomic ionization potentials, and Auger-decay times [24]. The excited electrons perform elastic and inelastic collisions until they lose their energy below a predefined cut-off. The elastic scattering is described with the screened Rutherford scattering cross section with modified Molier screening parameter [25]. For the inelastic scattering (impact ionizations and scattering on plasmons), the linear response theory is implemented with the single-pole approximation [26]. The calculated electron inelastic mean free paths, as well as combined photoabsorption attenuation lengths, are presented in the Appendix.

Electrons with energies below the cut-off, populating the valence and the bottom of the conduction band, are modelled with the Boltzmann collision integrals for the electron-electron and electron-phonon (electron-ion) scattering [27]. The electron-electron interaction is described with the relaxation-time approximation; in this work, the electron relaxation time is set to instantaneous thermalization, which ensures that the electronic distribution function adheres to the Fermi-Dirac distribution. The nonadiabatic electron-ion energy exchange is calculated with the dynamical coupling method [28].

The valence- and conduction band energy levels (band structure) evolve with the transferable tight-binding method [29–31]. The $sp^3d^5$-based PTBP density-functional tight-binding parametrization is used here, covering the pairwise interaction of all the elements involved [32]. The diagonalization of the electronic Hamiltonian produces the electronic energy levels (molecular orbitals) and the transient interatomic forces, dependent on the relative position of all the atoms in the simulation box [29]. The electronic distribution function (traced with the Boltzmann equation) directly affects the interatomic potential, enabling the description of the nonthermal melting and effects of bond breaking [31,33].

The atomic motion is traced with the molecular dynamics simulation, applying Martyna-Tuckerman's 4th-order algorithm with a time-step of 0.2 fs [34]. The simulation box contains 396 atoms, with the unit cell taken from Ref. [35], relaxed via the steepest descent algorithm, producing the equilibrium density of 2.73 g/cm$^3$. Then, the atomic velocities are initialized with the Maxwellian distribution at room temperature, allowing the material to thermalize before the arrival of the radiation pulse of 92 eV photon energy, 10 fs (FWHM) duration.

The illustrations of the atomic snapshots are prepared with the help of OVITO software [36].

## III. Results

We start by evaluating the phlogopite electronic density of states (DOS), since the electronic properties affect the atomic potential and will further help us to analyze the atomic behavior. The total and partial (projected) DOS in phlogopite are shown in Figure 1. They are evaluated on the 7x7x7 k-point Monkhorst-Pack grid in the entire supercell (396 atoms) [37]. The valence band is mainly formed by the *p*-states of oxygen atoms, whereas the bottom of the conduction band has significant contributions from K, Mg, Si, and Al atoms (H contribution is minor).

The calculated DOS qualitatively agrees with the previously reported DFT calculations [38]. The calculated band gap in the ideal crystal structure is 7.5 eV and shrinks to ~6.5 eV at room temperature, which also agrees reasonably well with various studies reporting the gap values between 4.8 eV and 6.9 eV [11,39]. This result validates applicability of the used PTBP tight binding parameterization to phlogopite [32].

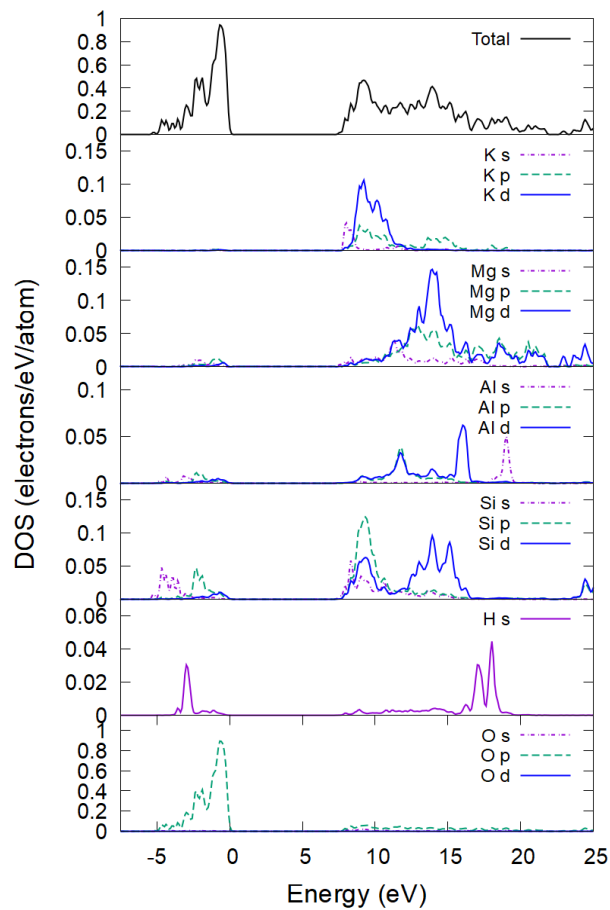

Figure 1. Total and partial electronic density of states in phlogopite counted from the Fermi level.

The irradiation of phlogopite was performed using an ultrashort laser pulse of 92 eV photon energy, 10 fs FWHM duration, and various deposited doses (energy densities) to identify the damage mechanisms and thresholds. The absorbed dose of 0.17±0.02 eV/atom (corresponding peak electronic temperature $T_e$~11,000 K, Figure 2) induces first defects: hydrogen migration, see Figure 2 showing atomic displacements of various species. The mean displacement of hydrogen grows continuously, whereas that of other elements saturates, indicating a stable lattice. Hydrogen diffuses, crossing the Mg layer, as seen in Figure 3. During this time, the electronic and atomic temperatures are still out of equilibrium. At this deposited dose, their equilibration requires tens of picoseconds, defined by the coupling parameter (see bottom panel of Figure 2), which is relatively small.

It is interesting to note that in a Born-Oppenheimer simulation (excluding electron-phonon coupling), the first damage occurs at a lower dose of ~0.15 eV/atom (~0.4% of valence electrons excited to the conduction band), and occurs via Mg and Si atom displacements into different planes, see Figure 4. This difference from the non-adiabatic simulation (cf. Figure 3) appears to be due to the electronic temperature being kept higher in the BO-simulation than in the non-BO one (Figure 4 vs. Figure 2): without the electron-phonon coupling, the electronic temperature stays constant.

The nonadiabatic (electron-phonon coupling) and nonthermal (changes of the interatomic potential due to electronic excitation) are competing in phlogopite. Electronic cooling due to electron-phonon coupling lowers the electronic temperature, returning the interatomic potential to the unexcited one, thereby precluding nonthermal damage in the Mg and Si subsystems and increasing the damage threshold. A similar effect of increasing the damage dose via electron-phonon coupling was observed in diamond [40].

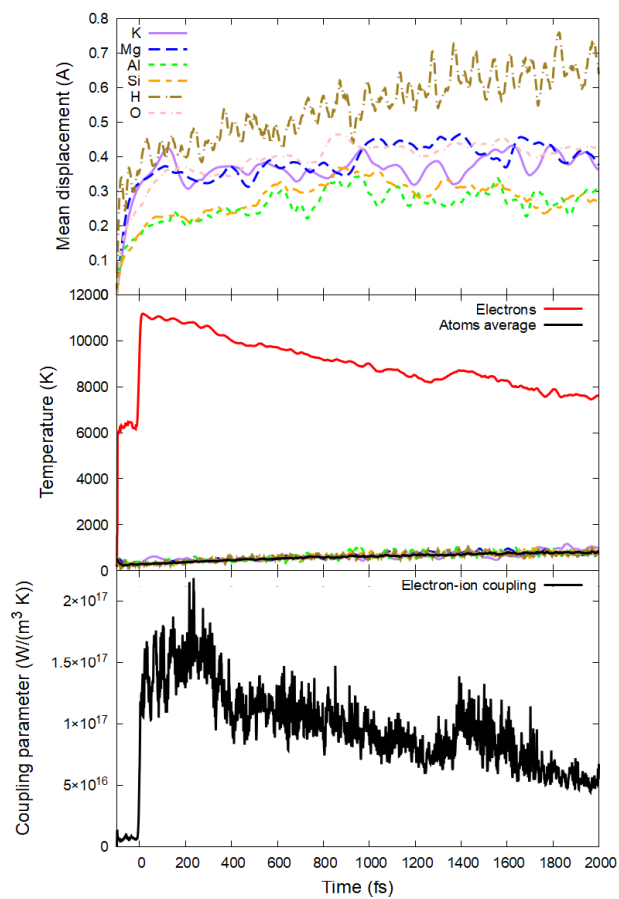

*Figure 2. (Top panel) Mean displacements of different elements; (middle panel) electronic and atomic temperature (total, and element-resolved); (bottom panel) electron-phonon coupling parameter, in phlogopite irradiated with 0.17 eV/atom, 92 eV photon energy, 10 fs FWHM duration.*

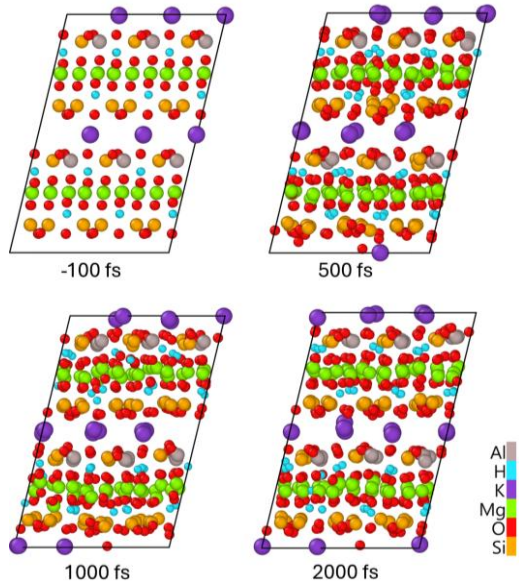

*Figure 3. Atomic snapshots of phlogopite irradiated with 0.17 eV/atom, 92 eV photon energy, 10 fs FWHM duration, simulated within nonadiabatic approximation (electron-phonon coupling included).*

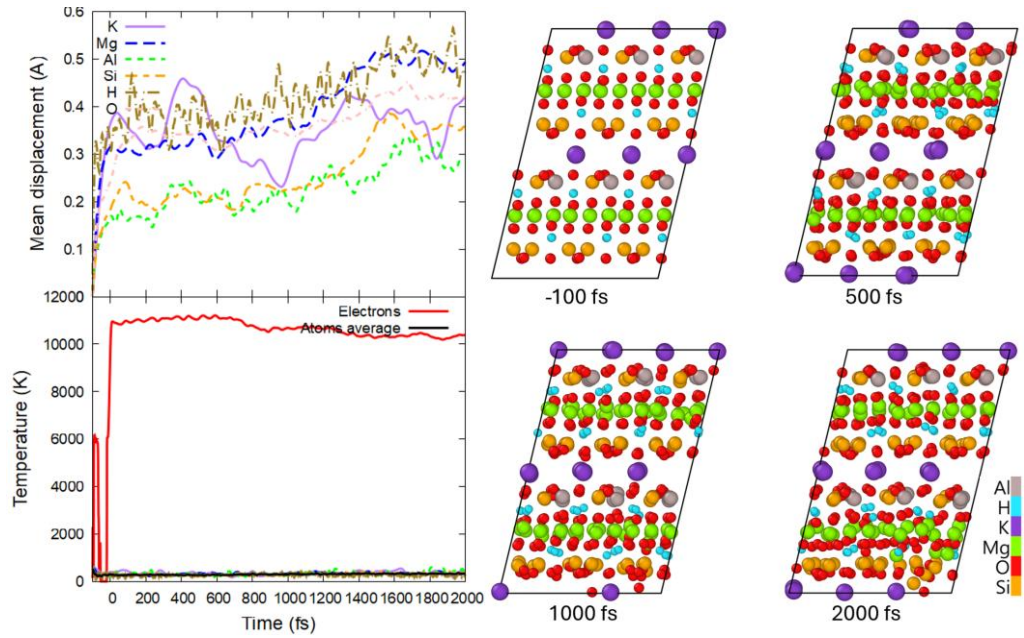

*Figure 4. (Top left panel) Mean displacements of different elements; (bottom left panel) electronic and atomic temperature (total, and element-resolved); (right panels) atomic snapshots of phlogopite irradiated with 0.15 eV/atom, 92 eV photon energy, 10 fs FWHM duration, simulated with BO approximation (excluding electron-phonon coupling).*

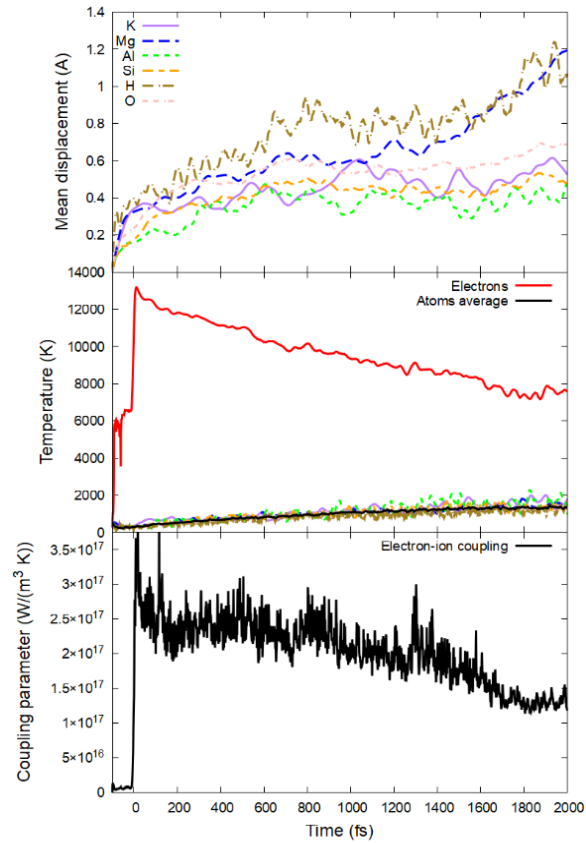

*Figure 5. (Top panel) Mean displacements of different elements; (middle panel) electronic and atomic temperature (total, and element-resolved); (bottom panel) electron-phonon coupling parameter, in phlogopite irradiated with 0.4 eV/atom, 92 eV photon energy, 10 fs FWHM duration.*

In the non-BO simulation, Mg atoms start to displace at the dose of ~0.4 eV/atom ($T_e$~13,000 K, see Figure 5). At such doses, the diffusion of the Mg atoms occurs within stable sublattices of other elements. In this state in phlogopite, Al, K, O, and Si form a solid lattice, whereas H and Mg are liquid-like. It suggests that around this deposited dose, a superionic-superionic phase transition takes place – from a state with liquid-like hydrogen, to a two-liquid-subsystems state.

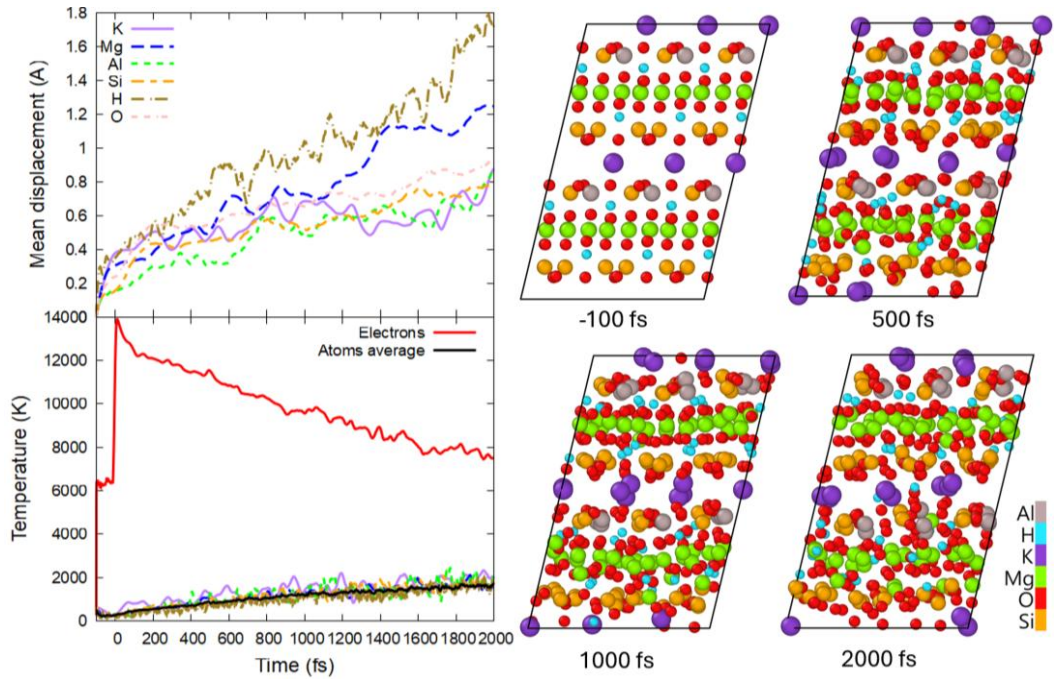

*Figure 6. (Top left panel) Mean displacements of different elements; (bottom left panel) electronic and atomic temperature (total, and element-resolved); (right panels) atomic snapshots in phlogopite irradiated with 0.5 eV/atom, 92 eV photon energy, 10 fs FWHM duration.*

With an increase of the dose to ~0.5 eV/atom ($T_e$~14,000 K), the entire atomic lattice destabilizes, disordering at a few-ps timescale, see Figure 6. Nonetheless, the complete melting of the system occurs with different sublattices disordering at different rates, with H and Mg subsystems disordering the fastest, whereas Al and K are the slowest. The phlogopite disorder is accompanied by the formation of defect energy levels inside the bandgap, see Figure 7.

Above the dose of ~0.9 eV/atom ($T_e$~16,000 K), the bandgap completely collapses (bottom panel in Figure 7), turning the material metallic (electronically conducting). Thus, another liquid nonequilibrium phase may be produced in phlogopite. We may conclude that the transient states in phlogopite may be superionic, semi-metallic, or metallic, depending on the deposited dose; completely disordered phlogopite is metallic.

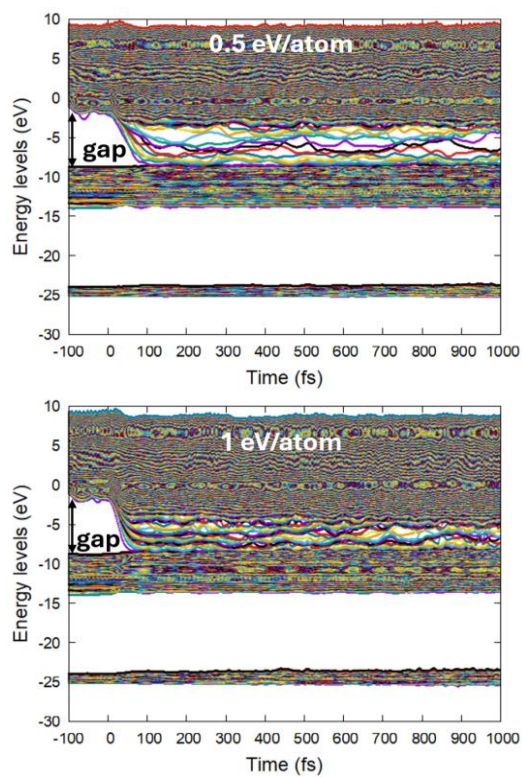

*Figure 7. Electronic energy levels (molecular orbitals) in phlogopite irradiated with the pulse of 92 eV photons, 10 fs FWHM duration, absorbed dose of 0.5 eV/atom (top panel) and 1 eV/atom (bottom panel). The initial band gap is indicated with the arrows.*

At even higher doses, nonthermal acceleration heats the atomic system at subpicosecond timescales, see an example of a 2 eV/atom dose in Figure 8 (peak electronic temperature ~20,000 K). The fastest element to accelerate is oxygen, followed closely by potassium and then other elements. Those elements acquire high kinetic temperatures within a few tens of femtoseconds. The increase in velocities leads to the rise in atomic displacements, as the system loses its stability and structure.

This selective atomic acceleration may be qualitatively explained by the electronic DOS structure (recall Figure 1): the electronic temperature increase smears out the Fermi-Dirac distribution function, removing electrons from the top of the valence band, promoting them to the bottom of the conduction band. The top of the valence band is predominantly formed by the oxygen *p*-states. As electrons are removed from there, the interatomic potential acting on oxygen atoms weakens and eventually turns more repulsive [41,42]. The bottom of the conduction band has large contributions from K (and Al, Si) states – as electrons are promoted into these states, the corresponding elements experience the modified interatomic potential and accelerate too.

As discussed in detail in Ref. [43], nonthermal atomic acceleration leads to an increase in the electron-phonon coupling parameter (which is proportional to the atomic temperature). The electron-phonon coupling is stronger at this irradiation dose than at lower ones (bottom panels in Figure 8 vs. Figure 2 and Figure 5). This self-amplifying process of nonthermal atomic acceleration, reinforcing atomic heating via electron-phonon coupling, leads to a complete atomic disorder on the scale of a few hundred femtoseconds.

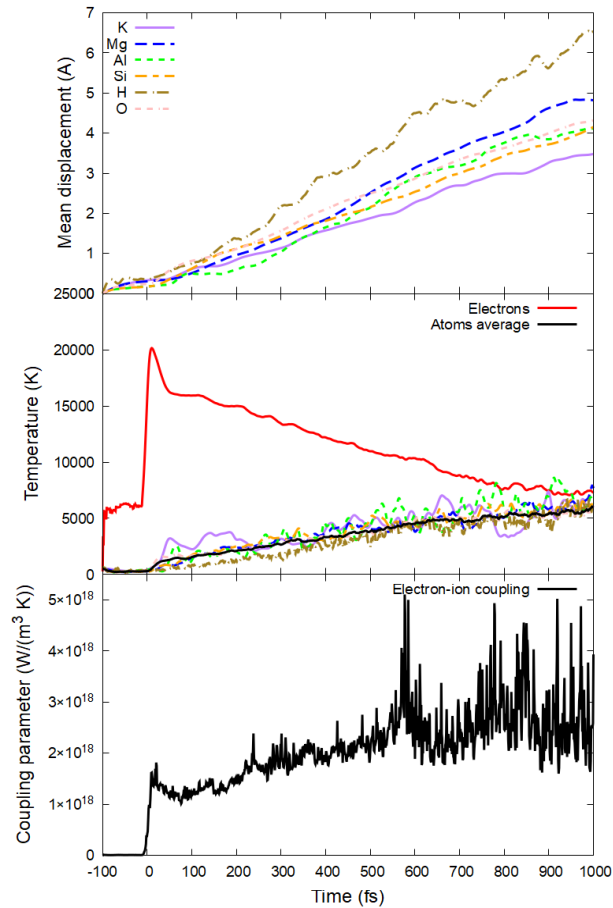

*Figure 8. (Top panel) Mean displacements of different elements; (middle panel) electronic and atomic temperature (total, and element-resolved); (bottom panel) electron-phonon coupling parameter, in phlogopite irradiated with 2 eV/atom, 92 eV photon energy, 10 fs FWHM duration.*

## IV. Discussion

The experiments on β- and γ-irradiation of phlogopite reported damage occurring at the dose on the order of 1000 kGy, at which significant hydrogen migration takes place, leading to material swelling [2,3]. Considering the density of phlogopite and its average atomic mass, this dose converts to ~0.2 eV/atom, which is remarkably close to the calculated here damage threshold of 0.17 eV/atom. In our calculations, this damage is also associated with hydrogen bond breaking, detachment, and diffusion. These results validate our simulation.

In experiments on irradiation of phlogopite with swift heavy ions in Refs. [44,45], the ion tracks were visualized with the help of chemical etching. The etched tracks had two distinct shapes: at lower ion energy loss, they were laterally triangular and discontinuous in depth, whereas with an increase in the energy loss of the projectile, the etch pits turned into a continuous hexagonal shape [44,45]. This indicates two different damage mechanisms occurring at different deposited doses. It was suggested that the hexagonal shape is associated with the damage in the $SiO_4$-tetrahedron in phlogopite, whereas the triangular shape is formed by the hydroxyl group. Although these results are not directly comparable to our simulations, qualitatively, they support our conclusions that damage in the hydrogen-containing sublattice forms at lower doses than in the silicon-containing one.

Finally, having calculated the damage thresholds for various processes in terms of the absorbed dose, we may estimate the corresponding thresholds in terms of the incident photon fluence. For this conversion, it is assumed that the incident pulse is normal to the surface of the sample, no reflection (typical for XUV or X-rays), no significant energy transport within the sample before the damage occurs, and emission of particles and energy from the surface. Under such assumptions, the damage threshold fluence is evaluated as $F = D\lambda n_{at}$, where D is the threshold dose, λ is the photon attenuation length at given photon energy (see Appendix), and $n_{at}$ is the atomic concentration in the sample. The damage threshold fluences in phlogopite for hydrogen migration, superionic H and Mg subsystems, and complete disorder are shown in Figure 9. These results may help to guide future experiments on phlogopite irradiation with X-ray pulses.

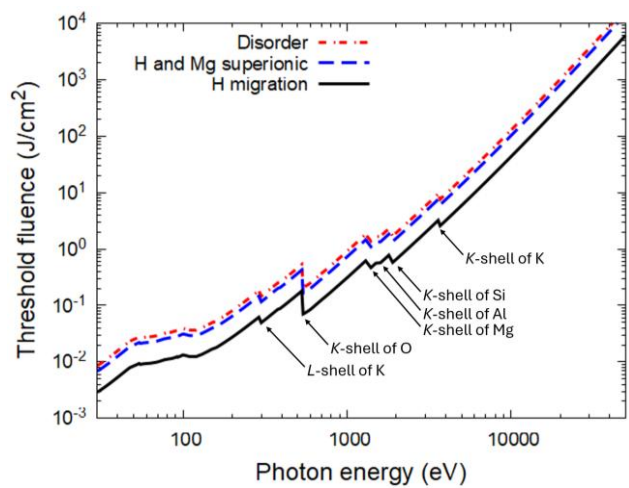

*Figure 9. Damage threshold fluences for in phlogopite for H migration, superionic H and Mg subsystems creation, and complete disorder. Contribution to photoabsorption from different shells of various elements, producing sudden jumps in the curves, and marked with arrows.*

## V. Conclusions

Phlogopite (KMg$_3$(AlSi$_3$O$_{10}$)(OH)$_2$) under electronic excitation is modelled with a combined code XTANT-3. Combining tight-binding molecular dynamics with transport Monte Carlo and Boltzmann equation enables the study of the nonequilibrium, nonthermal, and nonadiabatic effects in realistically complex materials.

Simulations predict that at the threshold dose of ~0.17 eV/atom (electronic temperature $T_e$~11,000 K), the first hydrogens detach and migrate. This estimated dose is in good agreement with the experimentally measured one, 1000 kGy=0.2 eV/atom. Curiously, the nonthermal and nonadiabatic effects in phlogopite are working in opposite directions: electron-phonon coupling, cooling the electronic system, precludes the nonthermal damage in the atomic system, increasing the damage threshold with respect to the Born-Oppenheimer simulation.

Increasing the deposited dose to ~0.4 eV/atom ($T_e$~13,000 K), Mg atoms start to diffuse like a liquid, whereas other sublattices remain stable, thereby forming a different superionic state. Above the dose of ~0.5 eV/atom ($T_e$~14,000 K), the entire atomic system disorders. At the doses of ~0.9 eV/atom ($T_e$~16,000 K), the electronic bandgap completely collapses, forming a liquid metallic state. At even higher doses (~2 eV/atom), nonthermal heating of the atomic system occurs at femtosecond timescales. This nonthermal effect is most pronounced in K and O elements, selectively accelerating them.

The simulation results suggest that a rich variety of transient states exist in irradiated phlogopite. Tuning the irradiation dose may enable production, tailoring, and studying of such states.

## VI. Conflicts of interest

There are no conflicts to declare.

## VII. Data and code availability

The code XTANT-3 is publicly available from [23].

## VIII. Acknowledgements

Computational resources were provided by the e-INFRA CZ project (ID:90254), supported by the Ministry of Education, Youth and Sports of the Czech Republic. The author thanks the financial support from the Czech Ministry of Education, Youth, and Sports (grant nr. LM2023068), and from the European Commission Horizon MSCA-SE Project MAMBA [HORIZON-MSCA-SE-2022 GAN 101131245].

## IX. Appendix

The total and partial photon attenuation lengths in the phlogopite, constructed from the EPICS2025 database [24] (for core shells), accounting for the stoichiometry of the material, and CDF-based cross section (for the valence band) are shown in Figure 10 (left panel) [26]. The electronic inelastic mean free paths in the same material, calculated with the linear response theory using the single-pole approximation, accounting for plasmon scattering [26], are shown in Figure 10 (right panel).

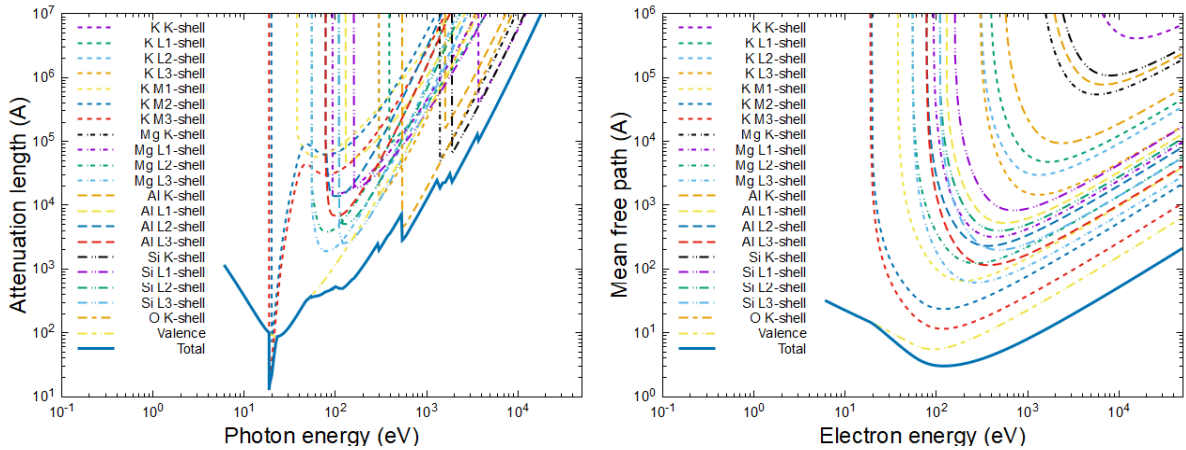

*Figure 10. Photon attenuation length (left panel) and electronic inelastic mean free path (right panel) in phlogopite. Total values are shown in solid lines. Partial ones, for each atomic shell of each element, are differently styled dashed.*

# X. References

[1] W.A. Deer, R.A. Howie, J. Zussman, An Introduction to the Rock-Forming Minerals, 3rd ed., Mineralogical Society of Great Britain and Ireland, Wirral, UK, 2013. https://doi.org/10.1180/DHZ.

[2] H. Wang, Y. Sun, J. Chu, X. Wang, M. Zhang, Intensive evaluation of radiation stability of phlogopite single crystals under high doses of γ-ray irradiation, RSC Adv. 9 (2019) 6199. https://doi.org/10.1039/C8RA08565J.

[3] H. Wang, C. Yang, X. Wang, J. Li, X. Su, K. Fang, J. Li, L. Jiang, An intensive exploration of the microstructural transformation undergone of phlogopite single-crystal film under electron beam (EB) irradiation at 0–1000 kGy: The influence of lattice stability on H-atom mobility, Ceram. Int. 49 (2023) 14445–14458. https://doi.org/10.1016/J.CERAMINT.2023.01.033.

[4] M. Lang, U.A. Glasmacher, B. Moine, C. Müller, R. Neumann, G.A. Wagner, Heavy-ion induced defects in phlogopite imaged by scanning force microscopy, Surf. Coatings Technol. 158–159 (2002) 439–443. https://doi.org/10.1016/S0257-8972(02)00270-0.

[5] A. Benedictus, P. Berendsen, A.M. Hagni, Quantitative characterisation of processed phlogopite ore from Silver City Dome District, Kansas, USA, by automated mineralogy, Miner. Eng. 21 (2008) 1083–1093. https://doi.org/10.1016/J.MINENG.2008.01.012.

[6] A.R. Cadore, R. De Oliveira, R. Longuinhos, V. de C. Teixeira, D.A. Nagaoka, V.T. Alvarenga, J. Ribeiro-Soares, K. Watanabe, T. Taniguchi, R.M. Paniago, A. Malachias, K. Krambrock, I.D. Barcelos, C.J.S. De Matos, Exploring the structural and optoelectronic properties of natural insulating phlogopite in van der Waals heterostructures, 2D Mater. 9 (2022) 035007. https://doi.org/10.1088/2053-1583/AC6CF4.

[7] H. Sreenivasan, P. Kinnunen, E.P. Heikkinen, M. Illikainen, Thermally treated phlogopite as magnesium-rich precursor for alkali activation purpose, Miner. Eng. 113 (2017) 47–54. https://doi.org/10.1016/J.MINENG.2017.08.003.

[8] A. Said, H. Hu, Y. Liu, Q. Zhang, J. Qu, Mechanochemical Activation of Phlogopite to Enhance its Capacity as Absorbent for the Removal of Heavy Metal Ions, Water. Air. Soil Pollut. 232 (2021) 1–10. https://doi.org/10.1007/S11270-020-04979-Z/FIGURES/8.

[9] M. Lang, F. Djurabekova, N. Medvedev, M. Toulemonde, C. Trautmann, Fundamental phenomena and applications of swift heavy ion irradiations, in: Compr. Nucl. Mater., Elsevier, 2020: pp. 485–516. https://doi.org/10.1016/B978-0-12-803581-8.11644-3.

[10] N. Medvedev, A.E. Volkov, R. Rymzhanov, F. Akhmetov, S. Gorbunov, R. Voronkov, P. Babaev, Frontiers, challenges, and solutions in modeling of swift heavy ion effects in materials, J. Appl. Phys. 133 (2023) 100701. https://doi.org/10.1063/5.0128774.

[11] T. Pucher, J. Hernandez-Ruiz, G. Tajuelo-Castilla, J.A. Ńgel Martín-Gago, C. Munuera, A. Castellanos-Gomez, Natural Layered Phlogopite Dielectric for Ultrathin Two-Dimensional Optoelectronics, ACS Nano 19 (2025). https://doi.org/10.1021/ACSNANO.5C09046.

[12] H. Palneedi, J.H. Park, D. Maurya, M. Peddigari, G.-T. Hwang, V. Annapureddy, J.-W. Kim, J.-J. Choi, B.-D. Hahn, S. Priya, K.J. Lee, J. Ryu, Laser Irradiation of Metal Oxide Films and Nanostructures: Applications and Advances, Adv. Mater. 30 (2018) 1705148. https://doi.org/10.1002/adma.201705148.

[13] B. Rethfeld, D.S. Ivanov, M.E. Garcia, S.I. Anisimov, Modelling ultrafast laser ablation, J. Phys. D. Appl. Phys. 50 (2017) 193001. https://doi.org/10.1088/1361-6463/50/19/193001.

[14] M. V Shugaev, C. Wu, O. Armbruster, A. Naghilou, N. Brouwer, D.S. Ivanov, T.J.Y. Derrien,

N.M. Bulgakova, W. Kautek, B. Rethfeld, L. V Zhigilei, Fundamentals of ultrafast laser-material interaction, MRS Bull. 41 (2016) 960–968. https://doi.org/10.1557/mrs.2016.274.

[15] F. Sarcan, N.J. Fairbairn, P. Zotev, T. Severs-Millard, D.J. Gillard, X. Wang, B. Conran, M. Heuken, A. Erol, A.I. Tartakovskii, T.F. Krauss, G.J. Hedley, Y. Wang, Understanding the impact of heavy ions and tailoring the optical properties of large-area monolayer WS2 using focused ion beam, Npj 2D Mater. Appl. 7 (2023) 1–7. https://doi.org/10.1038/S41699-023-00386-0;SUBJMETA.

[16] L.A. Smillie, M. Niihori, L. Rapp, B. Haberl, J.S. Williams, J.E. Bradby, C.J. Pickard, A. V. Rode, Exotic silicon phases synthesized through ultrashort laser-induced microexplosion: Characterization with Raman microspectroscopy, Phys. Rev. Mater. 4 (2020) 093803. https://doi.org/10.1103/PhysRevMaterials.4.093803.

[17] C.W. Siders, A. Cavalleri, K. Sokolowski-Tinten, C. Tóth, T. Guo, M. Kammler, M.H. von Hoegen, K.R. Wilson, D. von der Linde, C.P.J. Barty, Detection of nonthermal melting by ultrafast X-ray diffraction, Science 286 (1999) 1340–1342. https://doi.org/10.1126/science.286.5443.1340.

[18] A. Rousse, C. Rischel, S. Fourmaux, I. Uschmann, S. Sebban, G. Grillon, P. Balcou, E. Förster, J.P. Geindre, P. Audebert, J.C. Gauthier, D. Hulin, Non-thermal melting in semiconductors measured at femtosecond resolution., Nature 410 (2001) 65–8. https://doi.org/10.1038/35065045.

[19] R.O. Jones, Density functional theory: Its origins, rise to prominence, and future, Rev. Mod. Phys. 87 (2015) 897. https://doi.org/10.1103/RevModPhys.87.897.

[20] P. Verma, D.G. Truhlar, Status and Challenges of Density Functional Theory, Trends Chem. 2 (2020) 302–318. https://doi.org/10.1016/J.TRECHM.2020.02.005.

[21] T. Apostolova, E. Artacho, F. Cleri, M. Cotelo, M.L. Crespillo, F. Da Pieve, V. Dimitriou, F. Djurabekova, D.M. Duffy, G. García, M. García-Lechuga, B. Gu, T. Jarrin, E. Kaselouris, J. Kohanoff, G. Koundourakis, N. Koval, V. Lipp, L. Martin-Samos, N. Medvedev, A. Molina-Sánchez, D. Muñoz-Santiburcio, S.T. Murphy, K. Nordlund, E. Oliva, J. Olivares, N.A. Papadogiannis, A. Redondo-Cubero, A. Rivera de Mena, A.E. Sand, D. Sangalli, J. Siegel, A. V. Solov'yov, I.A. Solov'yov, J. Teunissen, E. Vázquez, A. V. Verkhovtsev, S. Viñals, M.D. Ynsa, Tools for investigating electronic excitation: experiment and multi-scale modelling, Universidad Politécnica de Madrid. Instituto de Fusión Nuclear Guillermo Velarde, Madrid, 2021. https://doi.org/10.20868/UPM.book.69109.

[22] C.L. Brooks, D.A. Case, S. Plimpton, B. Roux, D. Van Der Spoel, E. Tajkhorshid, Classical molecular dynamics, J. Chem. Phys. 154 (2021) 100401. https://doi.org/10.1063/5.0045455.

[23] N. Medvedev, XTANT-3 [Computer Software], (2023). https://doi.org/10.5281/zenodo.8392569.

[24] EPICS2025, (n.d.). https://nuclear.llnl.gov/EPICS/index.html (accessed September 25, 2025).

[25] T.M. Jenkins, W.R. Nelson, A. Rindi, Monte Carlo Transport of Electrons and Photons, Springer US, Boston, MA, 1988. https://doi.org/10.1007/978-1-4613-1059-4.

[26] N. Medvedev, F. Akhmetov, R.A. Rymzhanov, R. Voronkov, A.E. Volkov, Modeling time-resolved kinetics in solids induced by extreme electronic excitation, Adv. Theory Simulations 5 (2022) 2200091. https://doi.org/10.1002/ADTS.202200091.

[27] N. Medvedev, Electronic nonequilibrium effect in ultrafast-laser-irradiated solids, Phys.
14

Scr. 99 (2024) 015934. https://doi.org/10.1088/1402-4896/ad13df.

[28] N. Medvedev, I. Milov, Electron-phonon coupling in metals at high electronic temperatures, Phys. Rev. B 102 (2020) 064302. https://doi.org/10.1103/PhysRevB.102.064302.

[29] P. Koskinen, V. Mäkinen, Density-functional tight-binding for beginners, Comput. Mater. Sci. 47 (2009) 237–253. https://doi.org/10.1016/J.COMMATSCI.2009.07.013.

[30] H.O. Jeschke, M.E. Garcia, K.H. Bennemann, Microscopic analysis of the laser-induced femtosecond graphitization of diamond, Phys. Rev. B 60 (1999) R3701–R3704. https://doi.org/10.1103/PhysRevB.60.R3701.

[31] N. Medvedev, V. Tkachenko, V. Lipp, Z. Li, B. Ziaja, Various damage mechanisms in carbon and silicon materials under femtosecond x-ray irradiation, 4open 1 (2018) 3. https://doi.org/10.1051/fopen/2018003.

[32] M. Cui, K. Reuter, J.T. Margraf, Obtaining Robust Density Functional Tight-Binding Parameters for Solids across the Periodic Table, J. Chem. Theory Comput. 20 (2024) 5276–5290. https://doi.org/10.1021/acs.jctc.4c00228.

[33] H.O. Jeschke, M.E. Garcia, K.H. Bennemann, Theory for laser-induced ultrafast phase transitions in carbon, Appl. Phys. A 69 (1999) S49–S53. https://doi.org/10.1007/s003399900340.

[34] G.J. Martyna, M.E. Tuckerman, Symplectic reversible integrators: Predictor-corrector methods, J. Chem. Phys. 102 (1995) 8071–8077. https://doi.org/10.1063/1.469006.

[35] Materials Project, (n.d.). https://next-gen.materialsproject.org/ (accessed September 27, 2025).

[36] A. Stukowski, Visualization and analysis of atomistic simulation data with OVITO–the Open Visualization Tool, Model. Simul. Mater. Sci. Eng. 18 (2010) 15012. https://doi.org/10.1088/0965-0393/18/1/015012.

[37] H.J. Monkhorst, J.D. Pack, Special points for Brillouin-zone integrations, Phys. Rev. B 13 (1976) 5188–5192. https://doi.org/10.1103/PhysRevB.13.5188.

[38] G. Ulian, G. Valdrè, Crystal-chemical, vibrational and electronic properties of 1M-phlogopite K(Mg,Fe)3Si3AlO10(OH)2 from Density Functional Theory simulations, Appl. Clay Sci. 246 (2023) 107166. https://doi.org/10.1016/J.CLAY.2023.107166.

[39] G. Ulian, G. Valdrè, Crystallographic, electronic and vibrational properties of 2D silicate monolayers, J. Appl. Crystallogr. 58 (2025) 349. https://doi.org/10.1107/S1600576725000731.

[40] N. Medvedev, Z. Kuglerová, M. Makita, J. Chalupský, L. Juha, Damage threshold in pre-heated optical materials exposed to intense X-rays, Opt. Mater. Express 13 (2023) 808. https://doi.org/10.1364/OME.480936.

[41] N.S. Grigoryan, E.S. Zijlstra, M.E. Garcia, Electronic origin of bond softening and hardening in femtosecond-laser-excited magnesium, New J. Phys. 16 (2014) 13002. https://doi.org/10.1088/1367-2630/16/1/013002.

[42] S. Fukushima, N. Dasgupta, R.K. Kalia, A. Nakano, K. Shimamura, F. Shimojo, P. Vashishta, Photoinduced Phase Transition of Diamond: A Nonadiabatic Quantum Molecular Dynamics Study, J. Phys. Chem. Lett. 16 (2025) 9267–9272. https://doi.org/10.1021/acs.jpclett.5c01332.

[43] N. Medvedev, A.E. Volkov, Nonthermal acceleration of atoms as a mechanism of fast

lattice heating in ion tracks, J. Appl. Phys. 131 (2022) 225903. https://doi.org/10.1063/5.0095724.

[44] S.R. Hashemi-Nezhad, The triangular track contours in phlogopite mica detectors and discontinuity of the etchable damage, Nucl. Instruments Methods Phys. Res. Sect. B Beam Interact. with Mater. Atoms 142 (1998) 98–110. https://doi.org/10.1016/S0168-583X(98)00206-7.

[45] M.K. Lang, The Effect of Pressure on Ion Track Formation in Minerals, Combined Faculties for the Natural Science and for Mathematics of the Ruperto-Carola University of Heidelberg, 2004. https://archiv.ub.uni-heidelberg.de/volltextserver/5197/1/Dissertation_Lang_Maik.pdf (accessed September 27, 2025).